\documentclass[11pt]{article}
\usepackage{moriond,epsfig}
\usepackage{amsmath}
\usepackage{color} \usepackage{amssymb} \usepackage{graphicx}

\def\be{\begin{equation}}
\def\ee{\end{equation}}
\def\bea{\begin{eqnarray}}
\def\eea{\end{eqnarray}}

\newcommand{\DM}{\Omega_{CDM}h^2}
\newcommand{\DMW}{\Omega^{WMAP}_{CDM}h^2}
\newcommand{\gmu}{\delta a_{\mu}}

\newcommand{\stau}{\tilde{\tau}}

\newcommand{\neut}{\tilde{\chi}^0_1}

\newcommand{\DeltaO}{\Delta^{\Omega}}

\newcommand{\tanb}{\tan\beta}
\newcommand{\sigI}{1\sigma}
\newcommand{\sigII}{2\sigma}

\begin{document}
\vspace*{4cm}

\title{Natural SUSY Dark Matter: A Window on the GUT Scale}

\author{ J.~P.~Roberts }

\address{Institute of Theoretical Physics, Warsaw University, 69 Ho\.za,\\
Warsaw 00-681, Poland}

\maketitle\abstracts{ One of the key motivations for supersymmetry is
that it provides a natural candidate for dark matter. For a long time
the density of this candidate particle fell within cosmological bounds
across much of the SUSY parameter space. However with the precision
results of WMAP, it has become apparent that the majority of the SUSY
parameter space no longer fits the observed relic density. This has
given rise to claims that supersymmetry no longer provides a natural
explanation of dark matter. We address this claim by quantifying the
degree of fine-tuning required for the different dark matter
regions. We find that the dark matter regions vary widely in the
degree of tuning required. This degree of tuning can then be used to
provide valuable insights into the structure of SUSY breaking at the
GUT scale.}

\section{Introduction}

Supersymmetry at the TeV scale is one of the most compelling
candidates for physics beyond the Standard Model (SM). A primary
motivation for supersymmetry is that it removes the need to fine-tune
the bare Higgs mass. It also naturally provides a candidate for cold
dark matter. If we are to avoid fast proton decay, we must introduce a
symmetry that constrains the interactions of particles with their
supersymmetric partners. The most common form of this symmetry is
R-parity. This forbids the decay of a single superpartner into purely
SM matter. One result of this is that the lightest superparticle (LSP)
is absolutely stable. If SM matter and superparticles were in thermal
equilibrium in the early universe, the cooling universe would
leave behind a relic density of superparticles.

This has given rise to many claims that SUSY naturally accounts for
dark matter. However, having a candidate particle is one thing whereas
naturally accounting for the observed relic density is quite
another. In fact as WMAP has improved the constraints on the relic
density the regions of the SUSY parameter space that fit the observed
relic density have begun to look very slender. This has led to recent
claims that low energy supersymmetry requires significant fine-tuning
to fit the data that others claim it accounts for `naturally'.

One could ignore such a war of words over what is or is not
natural. However SUSY derives a significant portion of its motivation
from questions of tuning and naturalness. Therefore this question
deserves to be taken seriously. Here we present a quantititive study
of the fine-tuning required to access the different dark matter
regions of the Minimal Supersymmetric Standard Model (MSSM). We
discuss the implications of these different degrees of tuning for the
MSSM. Finally, we highlight how considerations of tuning allow us to
compare GUT scale models of SUSY breaking from LHC data.

\section{Fine-tuning and Dark Matter}

To quantitatively study fine-tuning we need a measure. The fine-tuning
required for electroweak symmetry breaking has a long history of
quantitative study. We follow Ellis and Olive \cite{Ellis:2001zk} in
using an analagous measure to study the fine-tuning of dark
matter\footnote{In contrast to Ellis and Olive, we take the total
fine-tuning of a point to be equal to the largest individual tuning
$\DeltaO=\text{max}(\DeltaO_a)$.}:
\begin{equation}
  \Delta_a^{\Omega}=\left|\frac{\partial
    \ln\left(\DM\right)}{\partial \ln\left(a \right)}\right|
\end{equation}
where $\{a\}$ are the free parameters of the theory. This provides a
measure of the sensitivity of the dark matter relic density to the
inputs. For example, if $\DeltaO_a=10$, a $1\%$ variation in $a$ would
result in a $10\%$ variation in $\DM$. \footnote{There many
alternative measures of fine-tuning that have been proposed in the
literature. We use this simple senstivity measure as it is easy to
understand, and allows for a straightforward comparison to the
electroweak fine-tuning price of SUSY models.}

\section{The Constrained Minimal Supersymmetric Standard Model (CMSSM)}

The MSSM is notorious for having over 100 free parameters. However,
many of these parameters are already constrained by experiment to be
zero. Furthermore, the parameters are only free if we leave the
mechanism of SUSY breaking entirely unspecified. In a realistic
theory, we would expect all the MSSM parameters to be set in terms of
a smaller set of more fundamental parameters. In the absence of a
specific theory of SUSY breaking we can still make progress. The most
frequesntly studied SUSY model is the CMSSM with the parameters:
\begin{equation}
  a_{CMSSM}\in \left\{ m_0,~m_{1/2},~A_0,~\tanb \text{ and
    sign}(\mu)\right\}
\end{equation}

Here $m_0$ is the common soft mass of all the scalar particles,
$m_{1/2}$ is the common mass for all the gauginos, $A_0$ sets the
third family trilinear couplings, $\tanb$ is the ratio of the two
Higgs vevs and $\mu$ is a bilinear Higgs mass term.

\begin{figure}[t]
  \begin{center}
  \psfig{figure=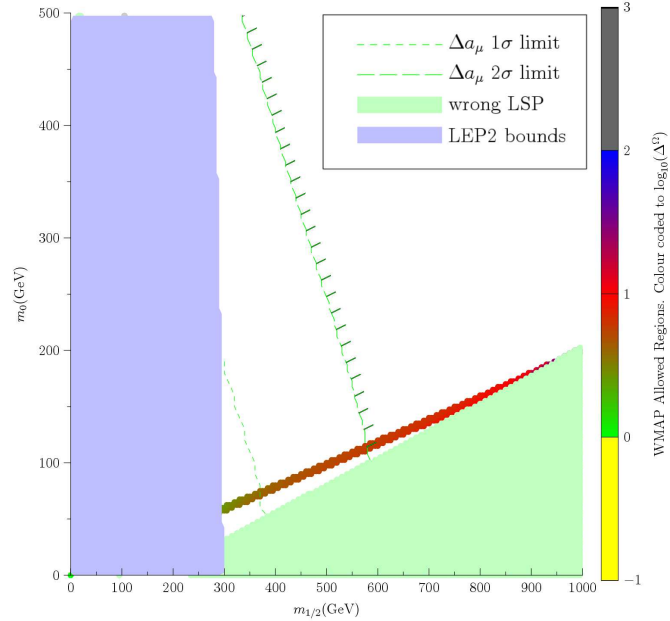,height=3.3in}
  \end{center}
  \caption{The $(m_{1/2},~m_0)$ plane of the CMSSM parameter space
    with $\tanb=10$, $A_0=0$ and $\mu$ positive. \label{fig:CMSSMt10}}
\end{figure}

In Fig.~\ref{fig:CMSSMt10} we show the $(m_{1/2},~m_0)$ plane of the
CMSSM parameter space for $\tanb=10$, $A_0=0$ and $\mu$ positive. Low
$m_0$ is ruled out (light green) as it results in a $\stau$ LSP. This
would result in a charged relic which would have been observed in
searches for anomalously heavy nuclei. Low $m_{1/2}$ is ruled out
(light blue) as this results in a light Higgs $m_h<111$~GeV. In the
remaining parameter space we plot the SUSY contribution to $(g-2)$ of
the muon, $\gmu$. We take the current observation of a deviation from
the Standard Model seriously and plot the region in which we agree
with the measurement at $\sigII$ (long dashed green lines) and $\sigI$
(short dashed green lines). It is clear that to explain the observed
value of $\gmu$ we require light soft SUSY masses and thus light
superpartners. Finally we plot the band that fits the observed relic
density of dark matter within $\sigII$. For every point that lies in
this band we calculate the fine-tuning and plot the point in a colour
that corresponds to the log-scale on the right of the plot.

The only dark matter region in Fig.~\ref{fig:CMSSMt10} lies alongside
the region in which the LSP is the $\stau$. Above this region the LSP
is the bino (the partner to the $U(1)$ gauge boson of the Standard
Model). Bino LSPs annihilate very weakly and normally give $\DM \gg
\DMW$. Thus, above the dark matter strip we have too much dark matter
and WMAP rules out the CMSSM. The remaining parameter space is very
slender. In the remaining parameter space, the $\stau$ and the
lightest neutralino, $\neut$, are very close in mass. This results in
a large number density of both particles in the early universe. The
resulting coannihilation of staus and neutralinos greatly enhances the
annihilation rate of SUSY matter, greatly lowering the resulting relic
density. This is very sensitive to the mass difference between the
stau and the lightest neutralino. Thus we would expect a
coannihilation region to be fine-tuned. However note that the band is
green at low $m_{1/2}$ and red at large $m_{1/2}$. This corresponds to
a tuning of $3-10$. This is a surprisingly small degree of
fine-tuning. The reason is that, for low $m_0$ and low $\tanb$, the
mass of the stau and the lightest neutralino are both primarily
dependent on $m_{1/2}$. Therefore the masses of the coannihilating
particles are coupled and the majority of the fine-tuning is removed.

\begin{figure}[t]
  \begin{center}
  \psfig{figure=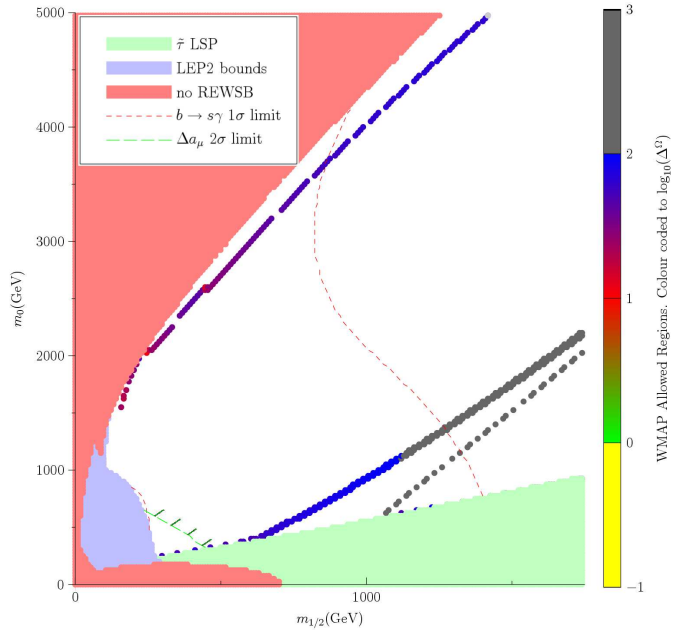,height=3.3in}
  \end{center}
  \caption{The $(m_{1/2},~m_0)$ plane of the CMSSM parameter space
    with $\tanb=50$, $A_0=0$ and $\mu$ positive. \label{fig:CMSSMt50}}
\end{figure}

This is a special case. In Fig.~\ref{fig:CMSSMt50} we show the
$(m_{1/2},~m_0)$ plane of the CMSSM with $\tanb=50$, $A_0=0$ and $\mu$
positive. We have also extended the range of $m_0$ and $m_{1/2}$. Many
of the bulk features remain the same. Low $m_0$ is still ruled out by
a $\stau$ LSP. Low $m_0$ and $m_{1/2}$ is ruled out by a light
Higgs. A new bound rules out large $m_0$ and low $m_{1/2}$ (light
red). Here the parameter space gives $\mu^2<0$ which is unphysical and
corresponds to a failure of radiative electroweak symmetry breaking
(REWSB).

The dark matter phenomenology is noticeably more complex. As before,
the LSP is bino across the majority of the parameter space and thus
mostly gives $\DM\gg\DMW$. The exceptions to this are marked by the
thin WMAP strips. Once again we have a coannihilation strip lying
along the side of the $\stau$ LSP region. In contrast to
Fig.~\ref{fig:CMSSMt10}, this band is plotted in purple, corresponding
to $\DeltaO\approx 50$.

This band is broken by two bands that go up in both $m_0$ and
$m_{1/2}$. These lie on either side of the line along which
$2m_{\neut}=m_A$ and the neutralinos annihilate via an s-channel
pseudoscalar Higgs boson. As could be expected, such a process
enormously enhances the annihilation of dark matter. We only fit the
dark matter relic density with {\it just enough} resonant
annihilation. This sounds like fine-tuning and indeed the lines are
mostly plotted in grey indicating $\DeltaO>100$.

Finally there is a dark matter band that runs alongside the region in
which $\mu^2<0$. Along the edge of this region we have low $\mu$ and
the higgsino fraction of the lightest neutralino increases. As bino
dark matter gives $\DM\gg\DMW$, and higgsino dark matter generally
gives $\DM\ll\DMW$, it is not surprising that somewhere in between we
manage to fit the observed dark matter relic density. However, the
requirement that the composition of the LSP include {\it just enough}
higgsino is an indication of fine-tuning and indeed the line is
plotted in purple and red indicating a tuning $\DeltaO\approx 30-60$.

\section{Breaking the Constraints}

\begin{figure}[t]
  \begin{center}
  \psfig{figure=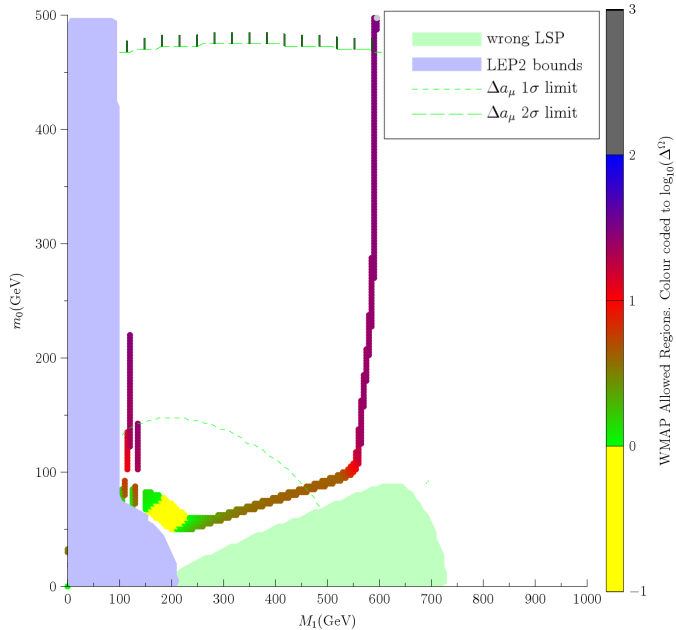,height=3.3in}
  \end{center}
  \caption{The $(M_1,~m_0)$ plane of the $CMSSM+M_i$ parameter
    space with $M_{2,3}=350$~GeV, $\tanb=10$, $A_0=0$ and $\mu$
    positive.\label{fig:Gauge}}
\end{figure}

We have shown the typical tunings of dark matter in the CMSSM, and
highlighted the problem that the bino LSP results in $\DM\gg\DMW$,
ruling out the majority of the CMSSM parameter space. However, there
are few compelling theoretical reasons to remain within the confines
of the CMSSM. Indeed there are many good reasons to relax a number of
the constraints. In previous work\cite{hep-ph/0603095} we study the
implications for fine-tuning of relaxing the constraint of universal
scalar masses and universal gaugino masses. In Fig.~\ref{fig:Gauge} we
consider a model in which we allow the gaugino masses to vary
independently of one another. Such a model has the parameters:
\begin{equation}
a_{CMSSM+M_i}\in\left\{ m_0,~M_1,~M_3,~M_3,~A_0,~\tanb \text{ and
sign}(\mu)\right\}
\end{equation}
where $M_{1,2,3}$ set the GUT scale soft SUSY breaking mass of the
superpartners to the $U(1)$, $SU(2)$ and $SU(3)$ gauge bosons
respectively. Such a break from gaugino mass universality can arise
naturally in string models \cite{hep-ph/0608135} and GUT models
\cite{0705.4219}.

In Fig.~\ref{fig:Gauge}, we show the $(M_1,m_0)$ plane of a model with
non-universal gaugino masses where we have fixed $M_{2,3}=350$~GeV,
$A_0=0$ and $\tanb=10$ with $\mu$ positive. As before there is a
region at low $m_0$ that is ruled out by a $\stau$ LSP. There is a
region ruled out at low $m_{1/2}$ due to light neutralinos and a
region ruled out at light $m_0$ and $m_{1/2}$ due to light sleptons.

The most notable feature is the explosion in the complexity of the
dark matter regions. Now, rather than the three dark matter regions of
the CMSSM, we have five distinct dark matter regions. Firstly, there
is the familiar bad along the edge of the $\stau$ LSP region due to
$\stau$-$\neut$ coannihilation. As before, this exhibits low
fine-tuning. This band is interrupted at $M_1\approx 570$~GeV. At
larger $M_1$ the neutralino is wino rather than bino. A wino LSP
generally gives $\DM\ll\DMW$. For $M_1<580$~GeV the neutralino is bino
so $\DM\gg\DMW$. Around $M_1\approx 570$~GeV the neutralino has {\it
just enough} bino and wino to fit the observed dark matter density. As
in the mixed bino/higgsino case, this requires a delicate balance and
thus the region exhibits a tuning $\DeltaO\approx 30$.

At low $M_1$ there are two distinct peaks at $M_1=110$~GeV and
$M_1=130$~GeV. These correspond to neutralino annihilation via an
s-channel $Z$ or light Higgs boson respectively.

Finally, there is a wide band that fits the observe dark matter relic
density at low $m_0$ and low $m_{1/2}$. It lies alongside the region
that is ruled out by LEP searches for light sleptons. In this band,
the sleptons are light enough to enhance the decay of neutralinos via
t-channel slepton exchange to the point where we suppress the dark
matter relic density enough to fit the observed data. This decay
process is remarkably insensitive to the precise value of the soft
masses. This translates to $\DeltaO<1$, corresponding to no
fine-tuning.

\begin{table}[t]
  \caption{The typical tunings for dark matter regions within the
    MSSM.\label{tab:DMTuning}}
  \vspace{0.4cm}
\begin{center}
  \begin{tabular}{|l|l|l|}
    \hline
    Region & Typical $\DeltaO$ \\
    \hline
    Mixed bino/wino & $\sim 30$\\
    Mixed bino/higgsino & $30-60$\\
    Mixed bino/wino/higgsino  & $4-60$\\
    Bulk region (t-channel $\tilde{f}$ exchange)& $<1$\\
    slepton coannihilation (low $M_1$, $m_0$)& $3-15$\\
    slepton coannihilation (large $M_1$, $m_0$, $\tanb$)& $\sim 50$\\
    $Z$-resonant annihilation & $\sim 10$\\
    $h^0$-resonant annihilation & $10-1000$\\
    $A^0$-resonant annihilation & $80-300$\\
    \hline
  \end{tabular}
\end{center}
\end{table}

Fig.~\ref{fig:Gauge} presents one example of the dark matter
phenomenology of the wider MSSM. By relaxing the constraints of the
CMSSM we can find the typical tunings of the different dark matter
regions that exist within the MSSM. We list these in
Table~\ref{tab:DMTuning}. Thus we conclude that each region has a
typical tuning, and that there remain regions of the MSSM that require
no tuning to accomodate the observed relic density.

\section{Conclusions: interpreting fine-tuning}

We must be careful in our interpretation of these results. Just
because an MSSM dark matter region exhibits significant fine-tuning,
does not mean that such a region will not be found at a future
collider. These tunings have been calculated with respect to the MSSM,
which is an {\it effective theory}. The MSSM does not specify the
mechanism of SUSY breaking, instead we parameterise our ignorance with
soft SUSY breaking masses. We expect that these masses should be set
by a deeper theory. Thus a region that is tuned in the MSSM may have a
very different tuning within a specific model of SUSY breaking.

This variation of tuning between models provides us with a useful
tool. If the LHC finds signals for new physics in the form of new
particles and large quantities of missing energy, many will interpret
this as a SUSY mass spectrum. There will be many different high energy
models that fit the data, and probably many models that will also fit
the observed dark matter relic density. However this raw mass spectrum
will do little to tell us what relations must obtain between high
energy parameters.

If we analyse the sensitivity of the dark matter relic density in such
a scheme we test precisely this dependence between high energy
parameters. For example, it is only because the $\stau$ mass and the
$\neut$ mass in the CMSSM coannihilation region are both dominantly
set by $m_{1/2}$ that such a region has low tuning. Therefore we would
have to favour such an explanation of an observed coannihilation
region than a model that set both masses independently.

After we identify the relations that mitigate the fine-tuning, we can
go on to make further predictions. Thus questions of fine-tuning can
help to narrow down the candidate explanations for a given
experimental signal. Having done this, novel predictions can be made
on the basis of hypothesised GUT scale relations between the soft
masses, and these can be tested in future experiments. Thus
fine-tuning and naturalness should allow us to analyse and compare GUT
scale physics using LHC energy data.

\section*{Acknowledgments}
The work of JPR was funded under the FP6 Marie Curie contract
MTKD-CT-2005-029466.

\end{document}